# Quantum Two-Way Protocol Beyond Superdense Coding: Joint Transfer of Data and Entanglement

KRISTIAN S. JENSEN[1] (Graduate Student Member, IEEE),
LORENZO VALENTINI[2] (Graduate Student Member, IEEE),
RENÉ B. CHRISTENSEN[1,3], MARCO CHIANI[2] (Fellow, IEEE),
AND PETAR POPOVSKI[1] (Fellow, IEEE)
[1]Department of Electronic Systems, Aalborg University, 9220 Aalborg, Denmark
[2]CNIT/National Laboratory of Wireless Communications, Department of Electrical, Electronic and Information Engineering,
University of Bologna, 40126 Bologna, Italy
[3]Department of Mathematical Sciences, Aalborg University, 9220 Aalborg, Denmark

Corresponding author: Kristian S. Jensen (e-mail: ksjen@es.aau.dk).

This work was supported in part by the Velux Foundation, Denmark, through the Villum Investigator Grant "WATER" and in part by the Ministero dell'Università e della Ricerca, Italy, through Project PRIN 2022 under Grant 2022JES5S2.

**ABSTRACT** In this article, we introduce a generalization of one-way superdense coding to two-way communication protocols for transmitting classical bits by using entangled quantum pairs. The proposed protocol jointly addresses the provision of entangled pairs and superdense coding, introducing an integrated approach for managing entanglement within the communication protocol. To assess the performance of the proposed protocol, we consider its data rate and resource usage, and we analyze this both in an ideal setting with no decoherence and in a more realistic setting where decoherence must be taken into account. In the ideal case, the proposal offers a 50% increase in both data rate and resource usage efficiency compared to conventional protocols. Even when decoherence is taken into consideration, the quantum protocol performs better as long as the decoherence time is not extremely short. Finally, we present the results of implementing the protocol in a computer simulation based on the NetSquid framework. We compare the simulation results with the theoretical values.

## I. INTRODUCTION

Rapid advancements in quantum technologies have ignited significant interest in harnessing the additional capabilities offered by quantum mechanics in communication protocols. One such capability is superdense coding [1], which enables the transmission of two classical bits of information by exchanging a single quantum bit (qubit) while sacrificing a previously exchanged pair of entangled qubits. Nevertheless, the excitement diminishes somewhat when considering the generation and distribution of entangled pairs, as, overall, users will need to exchange two qubits to transmit two classical bits of information. In this article, we show that the two-way communication setup is inherently capable of overcoming the drawback of presharing a large number of entangled qubits.

The simplest communication setting is one way, where only one of the users has data to be sent. If both users have to send data, this is called two-way communication [2]. Here, the users must coordinate their transmit/receive actions to comply with the capabilities of the channel. For instance, in half-duplex, the channel does not allow both users to transmit and receive at the same time. The coordination between the users is achieved by exchanging protocol signaling information, also called metadata. In this way, designing a two-way communication protocol becomes the problem of balancing transmission of metadata and actual data. A simple strategy is time-division duplexing (see, e.g., [3]), where time is divided into slots, and each user is assigned as either sender or receiver.

These considerations are valid for both classical and quantum channels, the main difference being the properties of the information carrier and potentially the capabilities of the users. For instance, photons can be used to transmit information in both classical and quantum settings. In the latter case,

4100408

information is not only carried by the presence or absence of a photon, but also in its quantum properties.

In this work, we consider the transmission of classical bits by means of a quantum channel. Similar to [4], we describe the channel conceptually by exchanging energy units. These energy units are an abstract notion that dictate who can transmit and who cannot. For instance, the protocol may guarantee that a user can transmit without collision as long as the user has not depleted its energy units. Once the user transmits the last energy unit, the role of sender will be passed on to the other user. In this model, users are assumed to be synchronized such that they agree on the starting time of each time slot. For each time slot, the sender will either stay silent (represented by "0") or transmit an energy unit (represented by "1"). In this way, the scheme can be divided into rounds with each round consisting of one participant transmitting a sequence of the form $0\ldots01$.

Of particular importance for our work is superdense coding. If we assume that the two users each hold one qubit of an Einstein–Podolsky–Rosen (EPR) pair [5], one user can perform local operations on its qubit to create an arbitrary maximally entangled state for the pair. Subsequently, this user sends its qubit to the other user. As EPR states are orthogonal, the receiving user can distinguish between the four possible states, thereby decoding two classical bits. By taking turns doing superdense coding, the users achieve a two-way protocol based on time-division duplexing.

For our purposes, we assume the source generation of EPR pairs [6]. This means that the sending user is responsible for creating the quantum state and distributing one qubit to the receiving user. In general, the generation and distribution of entanglement is a major downside of superdense coding since it will take up one time slot without transmitting any information bits.

Another challenge associated with using such pairs is their susceptibility to decoherence. The main contributor to decoherence in optical fiber is polarization mode dispersion (PMD) [7], [8]. This phenomenon is shown to lead to sudden death of entanglement in all but a restricted set known as a decoherence-free subspace (DFS) [9], [10]. In [7], it is shown how to reach a DFS. In addition, Shtaif et al. [11] examine the feasibility of nonlocal PMD compensation for entanglement distribution through optical fiber, and other methods for compensation involving channel adjustments have also been carried out [12]. In [13], the limit for distributing quantum states that decohere is established. In addition, efficient high-fidelity distribution of qubits through optical fiber and chains of repeater nodes has been proposed, with the former shown to achieve optimal fidelity [14], [15].

Decoherence implies that beyond a certain time from their creation, the entangled pairs no longer possess the necessary properties for superdense coding. Consequently, incorporating EPR pairs into the quantum framework imposes more stringent timing demands on communication protocols in real-world scenarios.

Furthermore, we assume that both nodes are capable of performing interaction-free measurements to detect the arrivals of qubits [16], [17]. In this way, users can detect the reception of qubits without disturbing the entanglement. This has been realized in practice using photons [18], [19], [20], as well as ultracold atoms [21]. Kwiat et al. [18] show that the fraction of measurements that are successfully interaction-free can be made arbitrarily close to 1, under the assumption that the system is lossless. This is achieved using a series of connected interferometers and beam splitters with specific reflectivities depending on the length $N$ of the series. Then, letting $N$ increase, the probability of detecting the photon goes to 1, while the probability of the photon getting absorbed goes to 0.

Based on the aforementioned observations, we ask the questions "*Is there a more efficient way to use superdense coding than simply using time-division duplexing?*" and "*Can such a procedure avoid preshared entanglement altogether?*" We answer these in the affirmative by presenting a two-way quantum communication scheme that mitigates the drawbacks associated with EPR generation and exchange by integrating them into the communication protocol. This allows us to increase information throughput per channel use compared to the conventional scheme. We explore this protocol in two settings: an ideal scenario where decoherence is not considered, and a more practical setting where the decoherence time is treated as a variable parameter.

The literature contains multiple examples of communication protocols between two users that can generate EPR pairs. As already mentioned, superdense coding [1] is the classical example. Another work is the so-called "ping-pong" protocol [22], which provides secure communication between two users and can be used for quantum key distribution. The security of this protocol, however, comes at a price. As we elaborate on in Section III, the throughput of the "ping-pong" protocol is significantly lower than our proposal. Yet another work considers the task of optimizing the entanglement rate over a network of chained quantum repeaters [23].

Another line of works focuses on communication protocols based on spatial superpositions [24], [25], [26]. These allow two classical bits to be transmitted in a single time slot, but rely on a specific setup of photon sources and beam splitters to generate the spatial superposition. In the current work, we consider a more general setting where the EPR pairs used are not restricted to a particular implementation or technology. In addition, our proposed protocol does not rely on a specific setup to generate entanglement (as is the case with spatial entanglement) and can hence be implemented in any quantum network.

In the purely classical setting, the idea of users passing energy units between each other has received some interest. In particular, works such as [27], [28], [29], [30], [31], and [32] consider an "energy harvest channel," and we also mention the use of a "token passing" channel in [33].

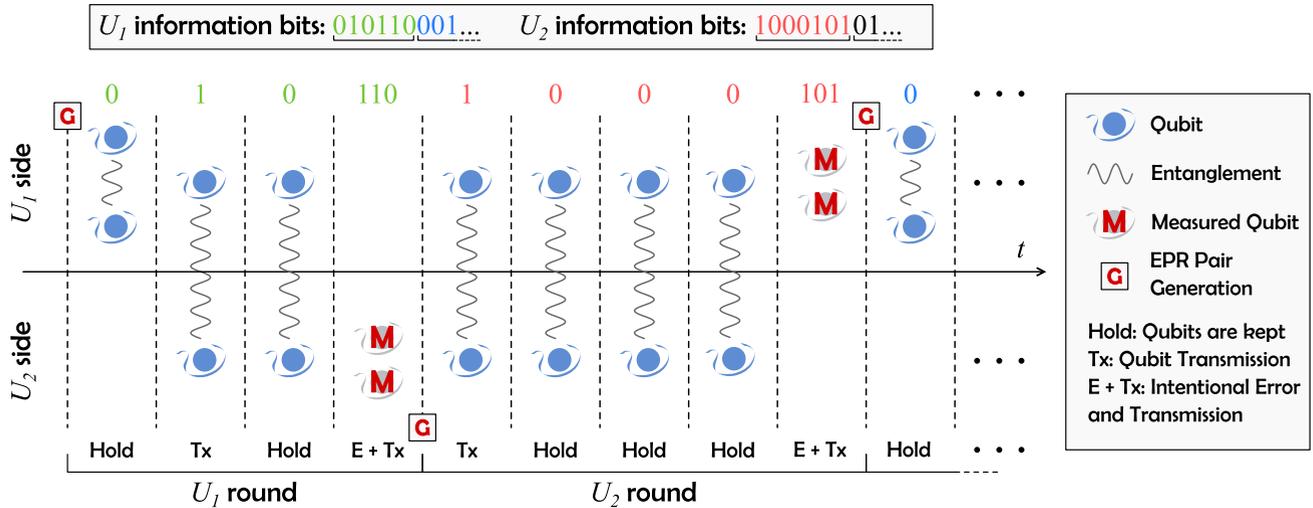

**FIGURE 1.** Proposed quantum two-way communication protocol to increase classical information throughput. Two users, $U_1$ and $U_2$, communicate bits by sending the qubits of an EPR pair through a quantum channel, in a slotted setting. This quantum transmission occurs each time a "1" appears as information bit to be transmitted. When the last qubit of the pair has to be transmitted, two extra information bits are superdense encoded in the EPR pair. This final burst of information bits provides an improvement compared to the classical scheme. The usage of entanglement sharing as a way to encode information provides an improvement compared to the one-way superdense code.

## II. PROTOCOL

We present a quantum two-way communication protocol, which exploits the quantum nature of the adopted carrier. In particular, as for superdense coding, we assume that classical information is exchanged by transmitting qubits. The communication scheme, illustrated in Fig. 1, works as follows. Two users, $U_1$ and $U_2$, each have a buffer containing information bits, which have to be transmitted to the other user. These bits are uniform as well as independent and identically distributed (i.i.d.).

Utilizing a classical approach, we define a "direct scheme" as a scheme where the presence or absence of a qubit (e.g., a photon) encodes a single classical bit of information. In this method, we just rely on qubit detection (e.g., with single-photon detectors). For each time slot, the sender will either stay silent ("0," no photon transmission) or transmit a qubit ("1," photon transmission). This method allows the overall scheme to be structured into distinct rounds, where each round involves a participant transmitting a sequence characterized by the pattern $0\ldots01$.

In contrast to the conventional direct scheme, our proposal operates under the assumption that users are equipped with the capability to generate and manipulate quantum states. Furthermore, users are expected to possess quantum memories for the storage of both generated and received qubits [34], [35]. In particular, we consider that a user generates an EPR pair and therefore holds two qubits. Hence, a round ends when two qubits are transmitted to the other side. Similarly to the direct scheme, the user does not transmit anything in the current slot when it has to communicate a "0," while it sends one of its qubits if a "1" has to be delivered. This would yield a sequence $0\ldots010\ldots01$ as in the classical sense. Up until now, this is equivalent to the direct scheme with two $0\ldots01$ sequences per round. The main difference introduced in the quantum scheme is the application of the following procedure when transmitting the last qubit of the pair. We apply an intentional Pauli error on the last qubit determined by the next two bits of the information buffer. In this way, we are able to transmit three information bits in a single slot: one bit due to the protocol (the "1," which triggers the transmission) and two bits due to superdense coding over an EPR pair. In fact, we can see superdense coding as piggybacking of classical data over a quantum carrier [36]. Note that a single quantum channel where both parties can communicate with each other, but not simultaneously (i.e., half-duplex system), is sufficient to permit the adoption of our protocol.

Regarding the metadata for coordinating the transmission, from Fig. 1, it is clear that the user that gets the two qubits also gets the right to transmit data. In terms of a protocol state machine, whenever the second qubit is transferred, the interpretation of the metadata is that the roles of sender and receiver are swapped and the other user gets the right to transmit.

To analyze protocols, we adopt two metrics: 1) the data rate $R$ and 2) the energy efficiency $E$. The data rate is defined as the number of information bits per time slot, and energy efficiency as the number of information bits per qubit transmitted. More formally, we define $R$ as

$$R = \lim_{n\to\infty} \frac{\sum_{i=1}^n \mathsf{B}_i}{\sum_{i=1}^n \mathsf{T}_i} \quad (1)$$

where $\mathsf{B}_i$ and $\mathsf{T}_i$ denote the number of bits and time slots of the $i$th round of transmissions, respectively. Then, due to the law of large numbers, we have

$$R = B/T$$

where $B = \mathbb{E}[\mathbf{B}]$ and $T = \mathbb{E}[\mathbf{T}]$. Similarly, the energy efficiency is defined as $E = \lim_{n\to\infty} \frac{1}{2n} \sum_{i=1}^{n} \mathbf{B}_i$ due to having two qubits sent in each round. As mentioned earlier, this reduces to

$$E = B/2.$$

A generic transmission round delivers an information bit sequence, which is composed of a stream of $\mathbf{K}_1$ zeros followed by a "1," concatenated with a second stream of $\mathbf{K}_2$ zeros followed by a "1" and by two extra bits. The latter two are encoded by superdense coding. The random variables $\mathbf{K}_i$, with $i = 1$ or $i = 2$, are i.i.d. geometrically distributed with success parameter one half due to equiprobability of bits "0" and "1." Their probability mass function is, therefore, $p_k = (1/2)^{k+1}$, with $k \geq 0$. A transmission round takes on average $T = 2\,\mathbb{E}[\mathbf{K}_i] + 2$ time slots and transmits on average $B = 2\,\mathbb{E}[\mathbf{K}_i] + 4$ information bits. Since $\mathbb{E}[\mathbf{K}_i] = 1$, we have that our proposal has a data rate of $R = 1.5$ and an energy efficiency of $E = 3$.

In comparison, the direct scheme can be analyzed in a similar fashion to reach $R = 1$ and $E = 2$, whereas the metrics for superdense coding are $R = 1$ and $E = 1$ (recall that we include entanglement distribution in the cost analysis). This shows that well-exploited quantum phenomena in communication schemes could provide significant advantages compared to classical ones.

We also note that the "ping-pong" protocol [22] is fairly easy to transform into a two-way protocol. Namely, the protocol includes periodic rounds in a "control mode." If users switch the roles of transmitter and receiver after each such control round, we obtain a secure two-way communication protocol. But even if we ignore the control rounds (where no information bits are exchanged), the performance metrics of this protocol are $R = 1/2$ and $E = 1/2$. As such, in settings where the quantum channel is trusted, our proposal gives significantly better performance.

## III. ACCOUNTING FOR DECOHERENCE

Let us now assume that, based on the particular technology implementation, the entanglement of the EPR pair can be maintained for a maximum of $c$ time slots. In this setting, for superdense coding to be possible, it is required that the number of time slots from entanglement generation to the second "1" being transmitted is at most $c$. For this reason, we propose a scheme variant of the quantum two-way communication protocol, which includes the following rules: the sender will postpone EPR generation to immediately before the first "1" appears in its input stream, and if this is followed by at least $m = c - 2$ zeros, the user will pad with an extra "1" (this operation is also referred to as bit stuffing [3]). In this way, the transmission is split into a string of zeros of unconstrained length and a string on the form $10 \cdots 01$ of length at most $c$. For example, having $m = 2$, the protocol does not perform bit stuffing to the sequence 00011, while it does for the sequences 001001 and 010001. Note that, despite the fact that even if $m$ zeros are already followed by a "1," the insertion of an extra "1" imposed by the scheme variant is essential to have a uniquely decodable sequence at the receiver. More precisely, if a round corresponds to string $0 \cdots 01s1$, where $s$ is a bit string, the receiver can distinguish two cases. If $s$ has length less than $m$, then the final "1" stems from the input buffer. Thus, the receiver obtains one bit in addition to the two from superdense coding. Otherwise, $s$ has length exactly $m$, and the final "1" stems from bit stuffing. In this case, the receiver only obtains information from superdense coding.

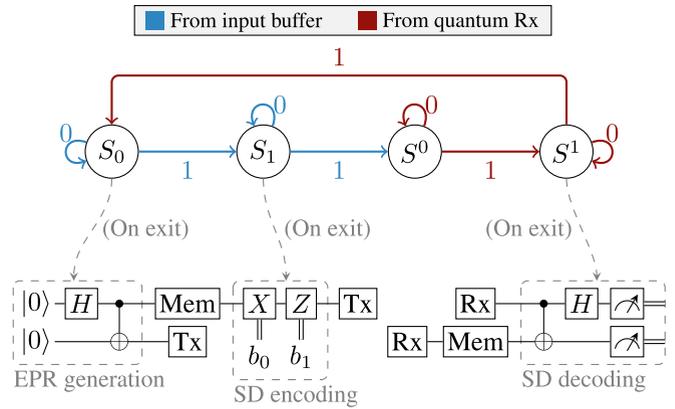

**FIGURE 2.** Graphical overview of the protocol from Section III. A detailed explanation on how to read the diagram is found in Section III-A.

### A. STATE MACHINE AND REPRESENTATION WITH QUANTUM CIRCUITS

In order to illustrate the scheme variant in the presence of decoherence, we give a graphical overview in Fig. 2. This figure contains a state machine where each state may be tied to a quantum circuit, and transitions may be determined by different inputs depending on the current state. While this is a nonstandard type of state machine, it should still aid in understanding the protocol. Note that Fig. 2 uses the assumption that bit stuffing has already been applied to the input buffer. If this were not the case, the state $S_1$ would need to be replaced by several states to ensure that transition to $S^0$ happens after at most $c$ time slots. In addition, state $S^1$ would need to be replaced in a similar fashion.

In greater detail, each user can be in one of four different states $S_0$, $S_1$, $S^0$, and $S^1$. The state $S_i$ represents the state of a sender having transmitted $i$ qubits in the current round. Similarly, $S^i$ represents the state of a received having received $i$ qubits. In states $S_0$ and $S_1$, the user will use the bits from its input buffer to determine state transitions. On the other hand, a user in state $S^0$ or $S^1$ will use its quantum receiver to determine state transitions with no reception interpreted as "0" and a received qubit interpreted as "1."

Upon leaving certain states, the user will have to execute some quantum circuit as indicated by the dashed lines in the diagram. For instance, once the sender transitions from state $S_0$ to $S_1$ (i.e., when it encounters the first "1" in its input bits), it will run the circuit labeled "EPR generation." One qubit is then stored in memory, and the other is transmitted immediately. Upon exiting from $S_1$ in a later time slot, the qubit in memory will be retrieved and input to the circuit labeled "SD encoding." Here, $b_0$ and $b_1$ represent the next two bits from the user's input buffer, which will be read and deleted when this circuit is executed.

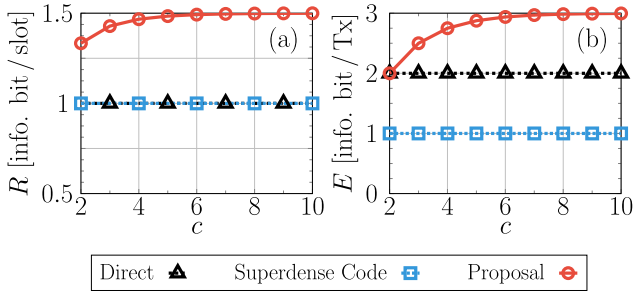

**FIGURE 3.** Performance of the proposed quantum protocol compared with two benchmarks: the direct version of such protocol and the superdense code. We vary the $c$ parameter that accounts for coherence time, reporting the data rate $R$ and the energy efficiency $E$. When decoherence is not too severe, the proposed scheme exhibits a 50% boost compared to the direct scheme.

### B. PERFORMANCE ANALYSIS

To derive the metrics of this variant scheme, we follow the same approach previously discussed. The random variable $K_1$ is exactly as before, but $K_2$ is restricted to be at most $m$. Thus, the probability mass function for $K_2$ is

$$p_i = \begin{cases} (1/2)^{i+1}, & 0 \leq i < m \\ (1/2)^i, & i = m \\ 0, & \text{elsewhere}. \end{cases}$$

Having that $\mathbb{E}[K_1] = 1$, one may derive

$$B = 4 + m\, p_m + \sum_{i=0}^{i-1}(k+1)\, p_i = 6 - \frac{1}{2^{m-1}}.$$

Similarly, we have that $T = 4 - (1/2)^m$. Putting $m = c - 2$, we obtain

$$R = 1 + \frac{1 - 1/2^{c-1}}{2 - 1/2^{c-1}} \qquad E = 3 - \frac{4}{2^c}. \qquad (2)$$

In Fig. 3, we plot the data rate $R$ and the energy efficiency $E$ of the proposed quantum scheme compared to the direct scheme, as well as the standard superdense-coding-based one. We see an uplift in performance already at $c = 2$, which translates to the EPR pairs remaining coherent just long enough for both qubits to be transmitted and measured. Regarding superdense coding, we point out that the advantage of this technique relies on the assumption that an infinite amount of preshared EPR pairs is available. Then, when accounting for this presharing procedure as overhead in the communication, the data rate $R$ degrades from 2 to 1 information bit per slot.

### IV. ACCOUNTING FOR DELAYS

In a practical implementation of our proposed scheme, there will be additional delays that affect the performance. For instance, the users will experience a propagation delay between the transmission and the arrival of each qubit. Within a transmission round, this is not an issue since the receiver can shift its time slots to match the delay, as illustrated in Fig. 4. This time shift must happen every time the roles of transmitter and receiver change during the protocol. In other words, each round will contain a fixed delay, leading to a penalty in transmission rate.

In addition, one may argue that the generation of an EPR pair is not instantaneous, and hence, this may also cause a delay. Once a round has started, the sender can peek ahead in the input buffer in order to know exactly when the next EPR pair is needed. Thus, in many cases, this generation can be performed "in the background," meaning that EPR pairs can be ready without delay. In general, however, this is a challenge with respect to rounds starting with a qubit transmission. Namely, the receiver cannot determine exactly when roles are changed and the next round will start. As such, it is not possible for the receiver to create a single EPR pair that is ready just in time.

To include these delays in the performance analysis, we introduce an additional parameter $\delta$ that denotes the required delay at the start of each round. Its unit of measure is "number of time slots," such that, e.g., $\delta = 1/2$ indicates that the incurred cost of EPR generation and propagation delay is equivalent to at most $1/2$ time slots.

With the parameter $\delta$, the rate $R$ from (1) is modified to

$$R = \lim_{n \to \infty} \frac{\sum_{i=1}^n B_i}{\sum_{i=1}^n (T_i + \delta)}$$

and similar arguments as before yield $R = B/(T + \delta)$. Note that the energy efficiency $E$ remains the same since the delay does not affect the number of qubit transmissions in a round. Applying this modified rate in the settings without and with decoherence, we obtain

$$R = 1 + \frac{1 - \delta/2}{2 + \delta/2} \qquad (3)$$

and

$$R = 1 + \frac{1 - 1/2^{c-1} - \delta/2}{2 - 1/2^{c-1} + \delta/2} \qquad (4)$$

respectively.

Note, however, that the impact of the delay can be made arbitrarily small by increasing the number of rounds between exchanging roles. More precisely, rather than switching between the transmitter and the receiver after every round, the users may do this every $N > 1$ rounds. In this case, the rate is obtained by replacing $\delta$ with $\delta/N$ in (3) and (4).

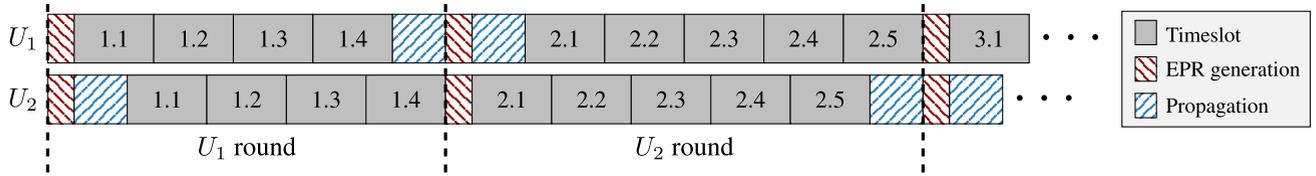

**FIGURE 4.** Graphical overview of the delays discussed in Section IV, with $U_1$ starting as the transmitter. Hatched regions denote delays and are divided into propagation delays and delays caused by EPR generation. Time slots have been numbered to emphasize the correspondence between them across users.

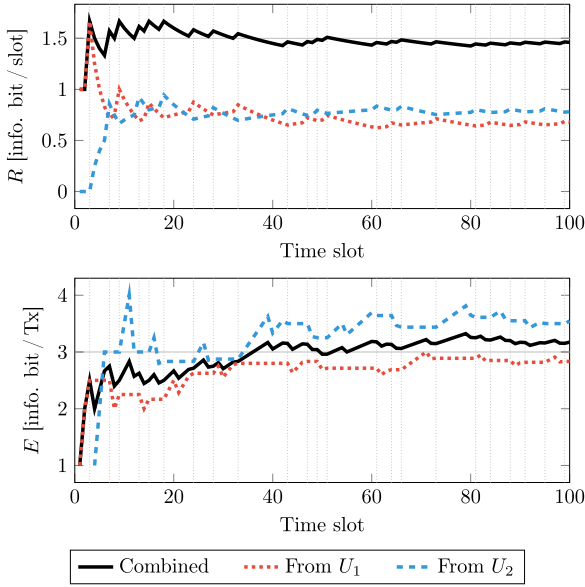

**FIGURE 5.** Simulated transmission rates $R$ and energy efficiencies $E$ for a single execution of the protocol described in Section II. Dotted vertical lines indicate the final time slot of a round.

## V. SIMULATION RESULTS

To support the analytical results of Sections II and III, we have performed simulations of the proposed protocols using the NetSquid framework [37].

For instance, Fig. 5 shows the transmission rate and energy efficiency as we track it across 100 time slots with error-free quantum channels and quantum memories. Upward spikes in the graphs for $R$ correspond to time slots where superdense coding is applied. These are also the time slots where the users swap their roles. As evident, the transmission rates settle around the theoretical 1.5 bits per time slot within the relatively short window of 100 time slots. Moving to the graphs for $E$, the interpretation is not quite as simple. In general, however, increases in the graph correspond to "long" rounds (i.e., rounds with many silent time slots), whereas decreases correspond to "short" rounds.

Of course, both $R$ and $E$ obtained in this way will be subject to random variation, so Fig. 6 shows the results of performing 1000 sample runs of the protocol. This illustrates that—as expected—averaging over more time slots yields transmission rates and energy efficiencies that are closer to the analytical asymptotic values. Note, however, that the transmission rates and energy efficiencies of all samples are significantly higher than 1, which is the comparable value for standard superdense coding.

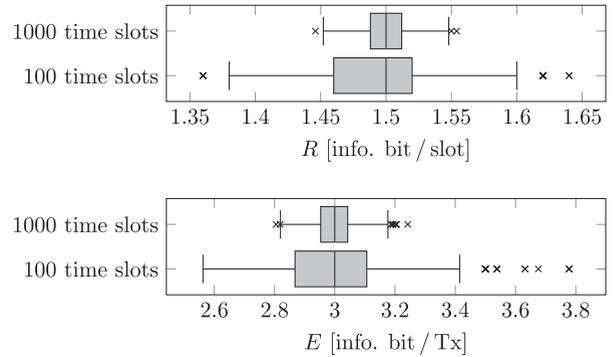

**FIGURE 6.** Boxplots summarizing the simulated transmission rates $R$ and energy efficiencies $E$ obtained at 100 and 1000 time slots, respectively. The results are based on 1000 sample runs of length 1000 time slots. The top boxplots use data from all 1000 time slots, whereas the bottom boxplots use data from the same executions, but only consider the first 100 time slots. An observation is considered to be an outlier if it is less than $q_1 - 1.5\text{IQR}$ or greater than $q_3 + 1.5\text{IQR}$, where $q_1$ and $q_3$ are the first and third quartiles, respectively, and $\text{IQR} = q_3 - q_1$ is the interquartile range. These outliers are marked with ×.

Moving to the case where decoherence takes effect, our simulation will use a more realistic noise model than the theoretical analysis in Section III. That is, rather than using a fixed coherence time, where EPR pairs are perfectly error-free for $c$ time slots and completely decohered in later time slots, we use NetSquid's memory noise model based on the relaxation times $T_1$ and $T_2$.[1] This does not change the analysis of the "raw" rates $R$ and $E$ of Section III, but does introduce a probability of error when decoding the bits sent via superdense coding. As such, one would, in practice, need to apply an error-correcting code, which would result in a lower effective rate. In this work, though, we refrain from performing such an analysis and report only the "raw" rates and the simulated error probabilities.

Fig. 7 shows the obtained transmission rates $R$ and energy efficiencies $E$ along with their theoretical values from (2). While the simulated transmission rates and energy efficiencies show some variation around their theoretical counterparts, there is generally agreement between the theoretical and simulated values.

---

[1]More precisely, we use `components.models.T1T2NoiseModel`.

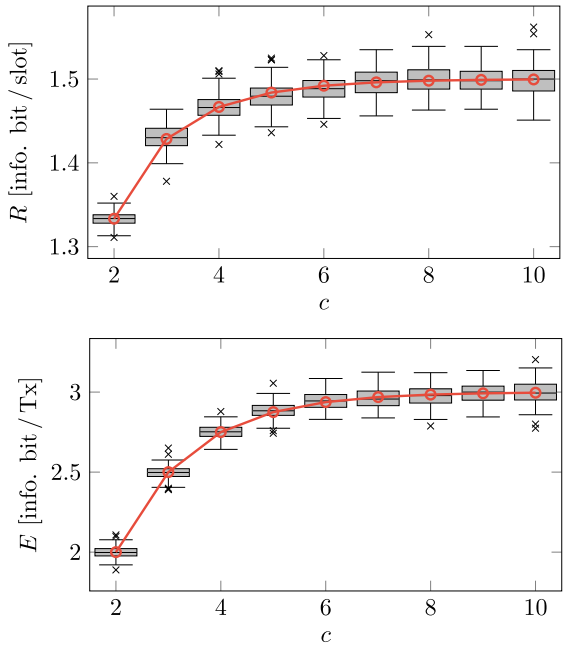

**FIGURE 7.** Simulated transmission rates *R* and energy efficiency *E* for the protocol described in Section III. For each value of *c*, 100 protocol executions were performed, each consisting of 1000 time slots. Outliers (as described in Fig. 6) are marked with ×. Points marked with "○" are the theoretical values from (2) (as also plotted in Fig. 3).

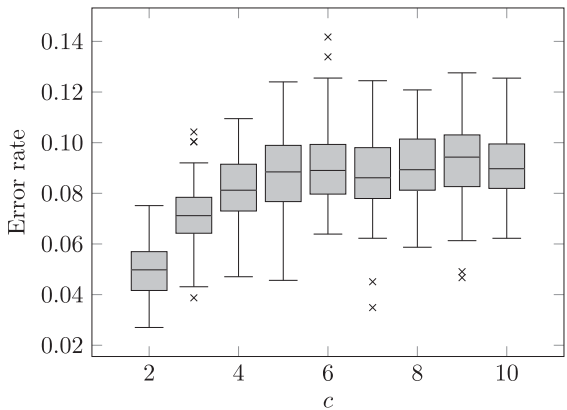

**FIGURE 8.** Simulated error rates in terms of erroneously decoded bits per bit transmitted using superdense coding. For each value of *c*, 100 protocol executions were performed, each consisting of 1000 time slots. Outliers (as described in Fig. 6) are marked with ×. Relaxation times were set to $T_1 = 20$ time slots and $T_2 = 18$ time slots.

In Fig. 8, the corresponding error rates of these simulation runs are presented. Here, we focus only on the bits transmitted using superdense coding since these are the only ones that can be erroneous. Therefore, the error rate expresses erroneously decoded bits per bit encoded with superdense coding. For instance, if $U_1$ transmits the sequence 10111, the final two bits are encoded using superdense coding. If $U_2$ erroneously decodes these as 10, we say that the error rate is 50%. In this way, the error rate can be used directly to choose appropriate error-correcting codes for these bits.

The relaxation times were set to $T_1 = 20$ time slots and $T_2 = 18$ time slots for these simulations. For this choice of parameters, we see that the error rates are significant even at small values of *c*. It is worth noting, however, that the error rate at $c = 2$ is identical to the error rate that one would see using superdense coding on its own. In other words, the magnitude of the error rate seems inherent to superdense coding with these relaxation times.

## VI. CONCLUSION

In this work, we have proposed a two-way communication scheme that addresses the main drawback of superdense coding, namely, the entanglement distribution. By weaving entanglement distribution into the protocol, we facilitate the transmission of information bits and the distribution of EPR pairs among participants. Furthermore, we provide a simple mathematical model to assess the protocol performance also accounting for quantum decoherence. In settings where EPR pairs are not too likely to lose coherence in the first few time slots, the proposal provides a performance uplift compared to the conventional schemes.

## VI. OPEN PROBLEMS

In this work, we have maintained a theoretical perspective on the derivation of our protocol. As such, we have limited the performance analysis to "ideal" circumstances where no errors occur. While not entirely realistic, it aids in explaining the main idea of the protocol in a simple manner without extra complications.

We leave open the problems of implementing supporting features such as protection against errors in the quantum channel or security against external attacks. This includes classical error correction for protecting the information bits within the quantum transmissions, as well as entanglement purification protocols for maintaining a sufficiently high fidelity of the EPR pairs [38]. For similar reasons, we leave the challenges of realizing our protocol in a practical setting as an open problem.

The work on error correction, however, is something we have already initiated. Specifically, we are currently working on extending this contribution with a theoretical analysis of potential error-correcting solutions to physical perturbations of the photons during their travel through optical fiber. While being a work in progress, this presently contains a proposal for combinations of classical error-correcting codes that, under certain assumptions on the error probability, show promising results on the asymptotic rate for certain classes of codes. This work, however, quickly expanded to the point where merging it with the current proposal seemed distracting from the aim of this article and would be more fitting as its own contribution.

Finally, although we have limited our research to performing all transmissions through optical fiber, there is currently an increasing focus on free-space quantum communications utilizing satellites [39], [40], [41]. This approach might circumvent some of the challenges presented by the application of optical fibers, but will require further examination.